\documentclass[10pt]{article}
\usepackage[T1]{fontenc}
\usepackage[latin1]{inputenc}
\usepackage{hyperref}
\usepackage{graphicx}
\usepackage{amsmath}
\usepackage{amssymb}

\newcommand{\drm}{{\rm d}}

\newcommand{\pa}{\partial}

\newcommand{\ug}{ \; = \; }

\newcommand{\bb}{\begin{equation}}
\newcommand{\ee}{\end{equation}}
\newcommand{\bega}{\begin{eqnarray}}
\newcommand{\ega}{\end{eqnarray}}
\newcommand{\begae}{\begin{eqnarray*}}
\newcommand{\egae}{\end{eqnarray*}}

\newcommand{\h}{\hspace*{4ex}}

\newcommand{\om}{\omega}

\newcommand{\cent}{\centerline}
\newcommand{\vs}{\vspace*}

\begin{document}

\baselineskip 0.5cm

\begin{center}

{\large {\bf Production of {\em Dynamic} Frozen Waves: Controlling shape, location (and speed) of diffraction-resistant beams } $^{\: (\dag)}$ } \footnotetext{$^{\: (\dag)}$ contact: mzamboni@decom.fee.unicamp.br}


\end{center}

\vs{0.2 cm}

\cent{T\'{a}rcio A. Vieira$^{\: 1,3}$, Marcos R.R. Gesualdi$^{\: 1}$, Michel Zamboni-Rached$^{\: 2,4}$, and
      Erasmo Recami$^{\: 2,5,6}$}

\vs{0.2 cm}

\centerline{{\em $^{\: 1}$ Universidade Federal do ABC, Santo Andr\'e, SP, Brazil.}}
\centerline{{\em $^{\: 2}$ University of Campinas, Campinas, SP, Brazil.}}
\centerline{{\em $^{\: 3}$ University of Sao Carlos, Sao Carlos, SP, Brazil.}}
\centerline{{\em $^{\: 4}$ University of Toronto, Toronto, Canada.}}
\centerline{{\em $^{\: 5}$ University of Bergamo, Bergamo, Italy.}}
\centerline{{\em $^{\: 6}$ INFN--Sezione di Milano, Milan, Italy.}}

\vs{0.5 cm}

{\bf Abstract  \ --} \ In recent times, we experimentally realized
a quite efficient modeling of the shape of diffraction-resistant
optical beams; thus generating for the first time the so-called
{\em Frozen Waves} (FW), whose longitudinal intensity pattern can
be arbitrarily chosen, within a prefixed space interval of the
propagation axis. Such waves possess a host of potential
applications: in medicine, biomedical optics, optical tweezers,
atom guiding, remote sensing, tractor beams, optical
communications or metrology, and other topics in photonic areas. \
In this work, we extend our theory of FWs ---which led to beams
endowed with a {\em static} envelope--- through a {\em dynamic}
modeling of the FWs, whose shape is now allowed to evolve in time
in a predetermined way. \ And we experimentally create such
dynamic FWs in Optics, via a computational holographic technique
and a spatial light modulator. \ Experimental results are here
presented for two cases of dynamic FWs, one of the zeroth and the
other of higher order, the last one being the most interesting,
consisting in a cylindrical surface of light whose geometry
changes in space and time.

\h PACS nos.: 42.25.Bs; 42.25.Fx; 41.20.Jb; 46.40.Cd; 41.85.-p; 46.40.Cd; 42.20.Ht; 42.40.Jv, 42.30Lr.


\vs{0.5 cm}

\h {\em 1. Introduction} --- The ``Frozen Waves"
(FWs)\cite{FW1,FW2,FW3} are very peculiar solutions to the wave
equations consisting in non-diffracting beams whose longitudinal
intensity pattern can be freely chosen within a prefixed space
interval $0 \leq z \leq L$ of the propagation axis. For such beams
even a certain degree of control on their transverse intensity
pattern is possible. This endows the FW-type beams with a great
deal of potential applications, since the concrete possibility of
modeling their spatial intensity shape {\em adds} to the
exceptional characteristic of all the ``Localized Waves", or
Non-Diffracting Waves (NDW), of resisting the effects of
diffraction\cite{19,20,JSTQE}.

\h The theoretical development of FWs can be found in
Refs.\cite{FW1,FW2,FW3,FW4}, besides in the above-quoted reviews
or books: We call it, briefly, the FW method. \ Synthesizing, a FW
is a superposition of co-propagating and equal-frequency {\em
Bessel beams} (of zeroth or higher order)
\begin{equation} \label{FW}
        \Psi(\rho,\phi,z,t) = \mathcal{N}_{\nu} e^{-i\omega t}\sum^{N}_{n=-N}A_{n}J_{\nu}(k_{\rho n}\rho)e^{ik_{zn}z}e^{i\nu\phi} \; ,
        \end{equation}
\noindent with $\mathcal{N}_{\nu} = 1/[J_{\nu}(.)]_{max}$, where
$[J_{\nu}(.)]_{max}$ is the maximum value of the Bessel function
of the first kind $J_{\nu}(.)$; $k_{\rho n}$ and $k_{z n}$ are
transverse and longitudinal wave numbers of the $n$-th Bessel beam
in the superposition, satisfying the known relation $k^{2}_{\rho
n} = \om^{2}/c^2 - k^{2}_{zn}$. \ The FW method requires the
longitudinal wavenumbers to be chosen so that \ $k_{zn} = Q + 2\pi
n /L$, quantity $Q$ being a constant to be selected, in the case
for instance of zero order FWs, according to the desired spot-size
$\Delta \rho_0$, besides for constructing solely (not evanescent)
beams {\em propagating} in the positive direction. (Let us recall
that a partial control on the beam transverse intensity too can be
obtained by a suitable choice of $Q$, and also by having recourse
to higher order Bessel beams).

\h An important point is that the FW method\cite{FW1,FW2,FW3,FW4}
shows that, once it has been arbitrarily chosen the desired
intensity longitudinal pattern $|F(z)|^2$ in the spatial interval
$0<z<L$, then the coefficients in superposition (\ref{FW}) have to
be \ $A_{n} = (1/L) \int^{L}_{0}F(z) e^{-i\frac{2\pi}{L}nz} \drm
z$. \

This is enough to obtain Diffraction-Resistant beams with the
required longitudinal intensity pattern, concentrated: \ \ (i) \
either along the propagation axis ($\rho=0$), when zero-order
($\nu=0$) Bessel beams are chosen in solution (\ref{FW}); it is
then possible to determine the spot radius of the resulting beam
from the parameter $Q$ via the relation ${Q} = (\omega^{2}/c^{2} -
2.4^{2}/\Delta \rho^{2}_{0})^{1/2}$; \ \ (ii) \ or over a
cylindrical surface, if a positive integer is adopted for $\nu$ in
(\ref{FW}); in which case, the cylinder radius can be
approximately evaluated through the equation \ $ \left. [ ({\drm}
/{\drm \rho}) J_{\nu}(\rho \sqrt{\omega^{2}/c^{2}-{Q}^{2}}) ]
\right|_{\rho=\rho_{0}} = 0 $.

\h It should be stressed that optical FWs have been concretely
generated via computational holography: see
Refs.\cite{Tarcio1,Tarcio2}; without forgetting that ultrasonic
FWs have been reproduced also in Acoustics, at least by simulated
experiments\cite{Prego1}. \ The experimental production of the FWs
opened the door to a great deal of applications in a number of
different areas: like medicine, biomedical optics, optical
tweezers, atom guiding, remote sensing, tractor beams, optical
communications or metrology, and so on.

\h Before going on, let us recall that FWs belong to the set of
the subluminal NDWs, corresponding to the very particular case of
zero peak-velocity\cite{LWs,NDWs}:  They possess, indeed, a {\em
static} envelope (within which only the carrier wave
propagates)\cite{FW1,FW2,FW3}. \ It appears convenient, however,
investigating, and constructing dynamic FWs, for all applications
in which a traveling FW appears to be desirable and useful. \ Aim
of this work is just extending the FW method towards a dynamic
modeling of the non-diffracting beams, whose shape can slightly
evolve in time in a predetermined way. That is, now we want our
resulting non-diffracting beams to be endowed with a spatial shape
which changes in time, as requested by the particular application
considered; \ and to produce experimentally such dynamic FWs in
Optics, via a computational holographic
technique\cite{Tarcio1,Tarcio2,Vasara,Arrizon1} in which computer
generated holograms of the FWs are created, as specified below,
and then reproduced by use of a spatial light modulator (SLM). \
[Such a resulting wave, that corresponds to an intensity pattern
no longer static, and slightly loses therefore its
monochromaticity, could not be called a beam, strictly
speaking...: However, if we consider the time evolution of its
shape much slower than the period associated to the relevant
frequency, our approach ends constituting an excellent
approximation].

\h In equation (\ref{FW}) we need inserting the time dependence,
first, of the coefficients $A_n$, and, second, of the parameter
$Q$ which enters the definition of the wavenumbers. \  Once it has
been chosen the desired longitudinal intensity $|F(z,t)|^2$, one
can exploit {\em the way} the said temporal variations are
implemented, so as to get the wished dynamical control on the
longitudinal shape of the FW (and at a certain extent on its
transverse shape too).

\h {\em 2. Theory of the ``Dynamic Frozen Waves"} --- Let us
consider the time-dependent solution
       \begin{equation}\label{FW_caminhante}
            \Psi(\rho,\phi,z,t) = \mathcal{N}_{\nu}\sum_{-N}^{N} A_{n}(t)
            J_{\nu} \left( k_{\rho n}(t)\, \rho \right)\,\exp(i\nu \phi)
            \exp \left( ik_{zn}(t) \, z \right)\,\exp(-i\omega t)
                   \end{equation}
\noindent where one may recall that any expression of the type \
$J_{\nu}(k_{\rho n}\rho)\exp(ik_{zn}z)\exp(-i\omega t)$, with
$k_{\rho n}^2 + k_{z n}^2 = \omega^2/c^2$, is a solution to the
wave equation \ $(\nabla^{2} - 1/c^{2} \partial_{t}^{2})\Psi = 0$.
\ As already mentioned, in the case of slow temporal variations,
we are able to obtain in the dynamical regime the longitudinal
pattern $F(z,t)$ and {\em either} the spot radius $\Delta
\rho_0(t)$ in the case of a zeroth-order FW; {\em or} the radius
$\rho_0(t)$ of the intensity cylindrical surface in the case of a
higher order FW.

\h Namely: To obtain a dynamic FW, instead of $F(z)$ we have to
consider for \ $0 \leq z \leq L$ \ a time-dependent function
$F(z,t)$.\ Then, considering superposition (\ref{FW_caminhante}),
one can determine the coefficients $A_{n}(t)$:
        \begin{equation} \label{An-t}
            A_{n}(t) = \frac{1}{L} \int^{L}_{0}F(z,t) e^{-i\frac{2\pi}{L}nz} \drm z \; ,
        \end{equation}
\noindent with the time-variation of the longitudinal or
transverse wave numbers \ $k_{\rho n}(t) = \sqrt{({2\pi} /
\lambda)^{2} - k_{zn}^{2}(t)}$ \ and \ $k_{zn}(t) \; = \; Q(t) +
2\pi n /L$ depending on $Q(t)$; which in its turn is linked either
to the spot size $\Delta \rho_0(t)$

\bb Q(t) = \sqrt{k^2 - ({2.4} / {\Delta \rho_0(t)})^2}
\label{Q1}\ee

in case of zero order FWs; \ or to the cylinder radius $\rho_0(t)$

\bb \left. [ ({\drm }/{\drm \rho}) J_{\nu}(\rho
\sqrt{\omega^{2}/c^{2}-{Q}^{2}(t)}) ] \right|_{\rho=\rho_{0}(t)}
\ug 0 \label{Q2}\ee

in case of higher order FWs.

\h We anticipated that a slow variation of $F(z,t)$ and $\Delta
\rho_0(t)$ (or $\rho_0(t)$), guarantees Eq.(\ref{FW_caminhante})
to be an approximate solution to the wave equation. \ This can be
shown by noting that for each term of the superposition
(\ref{FW_caminhante}) we have

\begin{equation} \label{derivada_temporal}
{\frac{\partial^2}{\partial t^2}}
 [A_{n}(t) J_{\nu}(k_{\rho n}(t)\rho) \exp(ik_{zn}(t)z) \exp(-i\omega t) ] \, \approx \,
A_{n}(t) J_{\nu}(k_{\rho n}(t)\rho) \exp(ik_{zn}(t)z)
{\frac{\partial^2}{\partial t^2}} \exp(-i\omega t)\, .
\end{equation}

\noindent The above approximation is valid due to the said slow
variation, which implies that $|\pa_t (A_n(t) J_{\nu}(k_{\rho
n}(t)\rho) e^{i k_{z n}(t) z}) | << | \om A_n(t) J_{\nu}(k_{\rho
n}(t)\rho) |  $. \ All this guarantees that
Eq.(\ref{FW_caminhante}) is a good approximate solution to the
wave equation.

\begin{figure}[!h]
                \centering
                \includegraphics[scale=0.45]{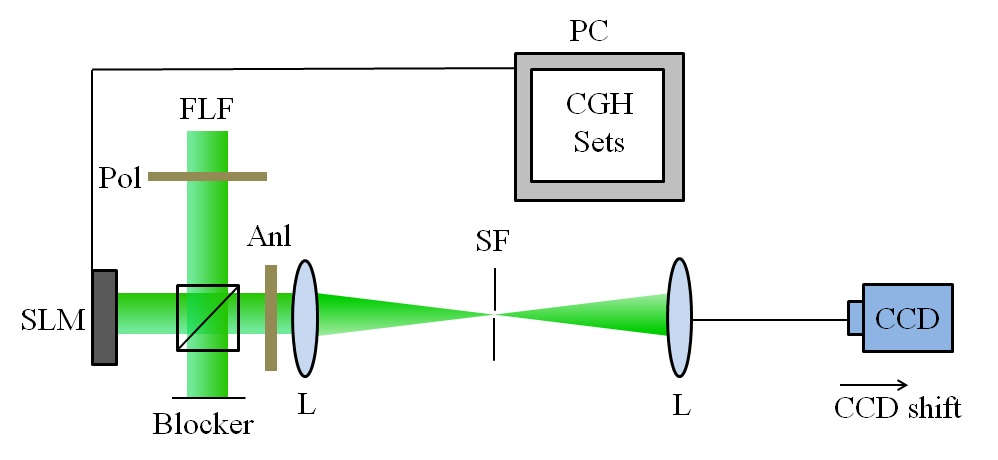}
                \caption{(color online) The experimental setup
used for generating a dynamic Frozen Wave is essentially composed
by an Argon ion laser beam, FLF being a beam collimator, Pol a
polarizer, BS a beam splitter, SLM a nematic Spatial Light
Modulator, and Anl an analyzer;  while the L's are lenses, SF is a
spatial filtering mask, and CCD is a camera for images/data
acquisition. }
                \label{arranjo}
        \end{figure}

\h {\em 3. Generation of Dynamic Frozen Waves {\rm (DFW)\/}:
Experiments and results} --- As already mentioned, the
experimental holographic method for the generation of {\em static}
FWs can be found in Refs.\cite{Tarcio1,Tarcio2}. \ Our present
setup is shown in Fig.\ref{arranjo}. It is composed by an Argon
ion laser (of 514,5 nm), which is expanded and collimated (Exp) in
a nematic SLM device (LC R1080 SLM, HoloEye Photonics AG). We
perform the amplitude modulation of a computer generated hologram
(CGH) on the SLM by polarizer Pol and analyzer Anl, with angles
$0^{\rm o}$ and $90^{\rm o}$, respectively, measured with regard
to the input SLM axis. The 4-f spatial {\em filtering}
system\cite{Arrizon1,Tarcio1,Tarcio2} is used for the FW
generation, the SLM (CGH) being placed at the input plane (focus
of the first lens); while a spatial {\em filtering} mask (SF, {\em
band-pass circular pupil\/}), located at the Fourier plane, {\em
selects and transmits} the light from the initial spectrum,
generating the encoded signal at the output plane of the setup. As
a result, we get a propagation of the FW, whose intensity is
registered with a CCD camera that can be displaced all along the
distance $0 < z <L$.

\h Specifically, in the case of DFWs we discretized the relevant
time interval (of 10 s, for instance) into ${m}$ parts; and we
obtained, for each  value $\tau_m$ of time, the coefficients of
the series (\ref{An-t}), and consequently the complex function
$\Psi(\rho,z,t=\tau_{m})$ describing the FW at that instant of
time. \ On knowing the field value at each time instant, $m$
computer generated holograms (CGHs) are created. \ Then, by using
a Spatial Light Modulator (SLM) linked to a microcomputer, the
holograms are reproduced at a specific rate on the modulator
display to get a reconstruction of the DFW: That is, of position
and longitudinal (and transverse) shape, varying with time, of the
dynamic frozen wave (see Fig.\ref{arranjo}). The LC-R1080
modulator possesses a rate of 70 Hz, enough for producing 15
frames per second.

\begin{figure}[!h]
                \centering
                \includegraphics[scale=0.3]{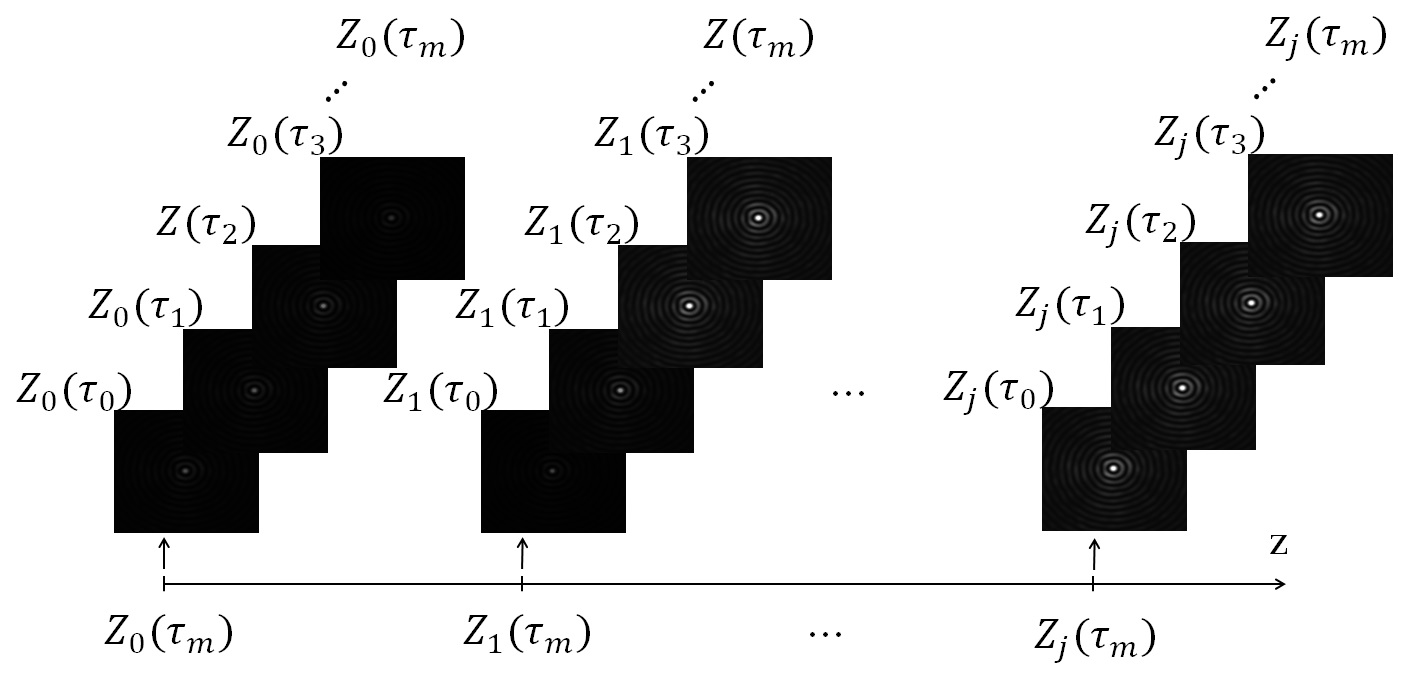}
                \caption{Scheme of the experimental data acquisition. The $z$-axis is divided into $j$ parts $Z_{j}$;  the data acquisition is performed leaving the CCD camera in the fixed position $z_{0} = 0$, where all frames $\tau_{m}$ reproduced by the SLM are captured. The process is repeated in the subsequent positions till the $j$-th one. }
                \label{analise}
        \end{figure}

\h The $z$-axis too is divided into $j$ parts $Z_{j}$. The data
acquisition is then performed leaving the CCD camera in the fixed
position $z_{0} = 0$ (the beginning of the propagation axis),
where all frames $\tau_{m}$ reproduced by the SLM are captured;
the process is repeated in the subsequent positions till the
$j$-th one, as in Fig.\ref{analise}.

\h 3.1. First example (for a zeroth-order DFW): -- Let us
consider, in the space interval $0<z<L$, an exponentially rising
barrier of the type
\begin{equation} \label{exponRisingDFW}
               F(z,t)= \left\{ \begin{array}{rl}
                                      \exp(K zt) , &  \text{for} ~~ z_{i} \leq z \leq z_{f} \\
                                       0, &  ~~~\text{elsewhere}
                \end{array}
                \right.
\end{equation}
\noindent with \ $z_{i} = z_{0} + vt$ \ and \ $z_{f} = \Delta D + z_{0} - vt$, \ quantity $\Delta D = z_{f} - z_{i}$ being the width of the considered DFW at the initial time instant, and $K^{-1} = 1$ cm s. \ The coefficients of our series, at each instant $t$ of time, are given by Eq.(\ref{An-t}) when inserting into it the function (\ref{exponRisingDFW}). \ Afterwards, the FW behavior in the considered time interval (e.g., $\Delta t = 10$ s) is obtained by putting such values $A_{n}(t)$ into solution (\ref{FW_caminhante}).

 \h Let us adopt the values: \ ${Q} = 0.99997 \,
\omega/c$, and $L = 40\;$cm, with $N = 20$ superpositions; \ a
step width $\Delta D = 4\;$cm; \ initial position $z_{0} =
0.18\;$cm, and finally a constant $v = 1.3\;$cm/s. \ The initial
and final positions move with velocity $-v$ and $v$, respectively.
\ {\em We then get} a step profile which, as time elapses,
modifies into the longitudinal exponential ramp expressed by
Eq.(\ref{exponRisingDFW}). \ More specifically, let us divide the
time interval $\Delta t = 10\;$s into $m=150$ parts: \ From
Eqs.(\ref{exponRisingDFW}) and (\ref{FW_caminhante}) we obtain
$\Psi(\rho,z,t=\tau_{m})$, and for each time value $\tau_{m}$ an
amplitude hologram is generated, with a total of 150 holograms
(frames).

\begin{figure}[!ht]
                \centering
                \includegraphics[scale=0.3]{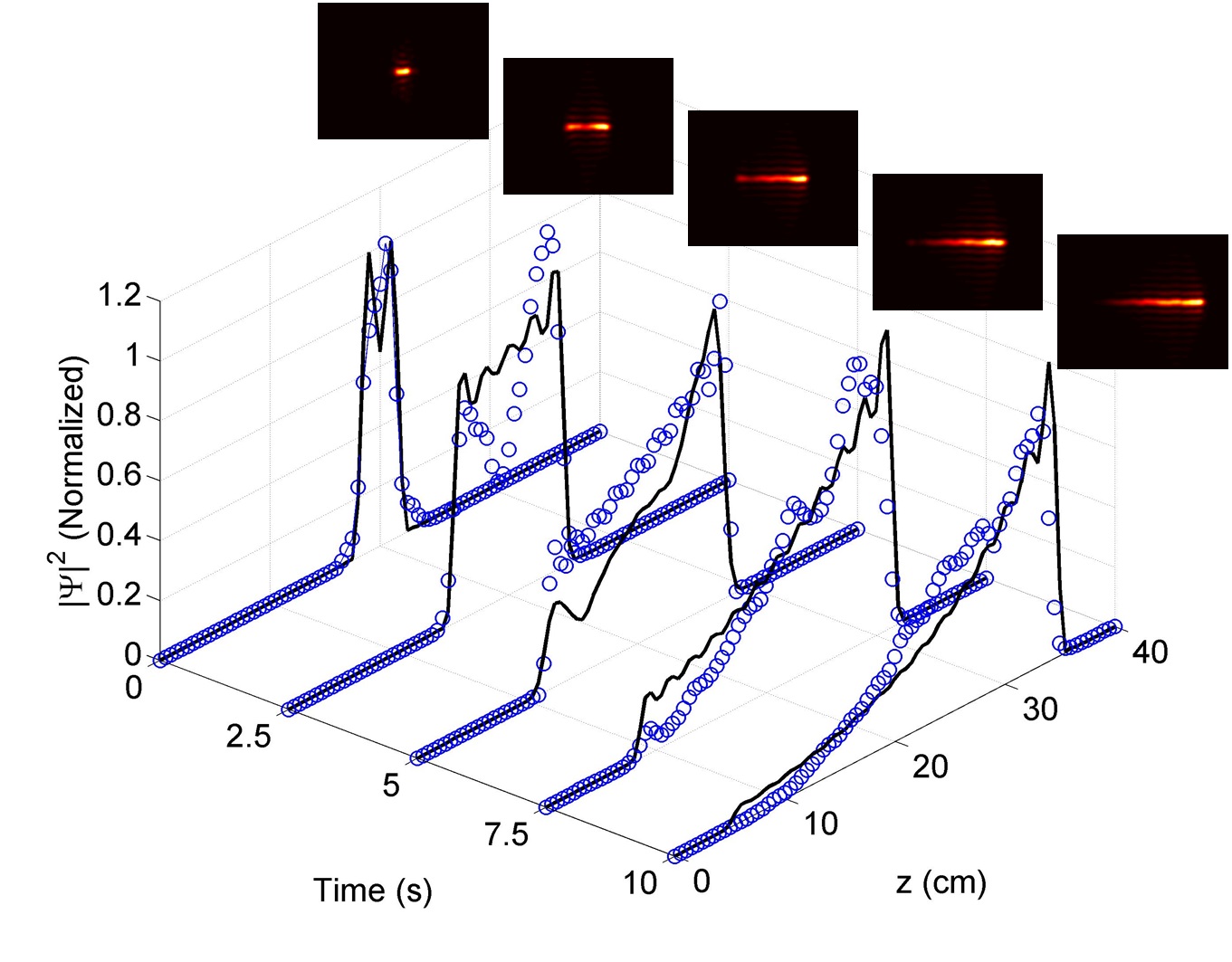}
                \caption{(color online) Time evolution (for $\Delta t = 10$ s) of the longitudinal intensity pattern for the FW
 corresponing to the longitudinal intensity profile given by an exponentially rising ramp like (\ref{exponRisingDFW}).  \  In the insets, the corresponding orthogonal projection are shown. See also Media-1 online.}
                \label{Tarcio-degrau_para_exp_com_proje}
        \end{figure}

\h In the present case, the $z$-axis was divided into 81 parts,
and 143 frames were captured at each position $Z(j)$.  This
implies an error in the temporal reconstruction of the holograms
(error due to the intrinsic limitations of the microcomputer
hardware and of the software itself: think of the time spent by
Matlab between command and execution) that can be evaluated to be
of about 7\%.
In Figs.\ref{Tarcio-degrau_para_exp_com_proje} the
longitudinal intensity profile for different time instants is
shown, together with its corresponding orthogonal projections
(while online, in ``Media-1" the 3-D profile is shown in a dynamic
way).

\h 3.2. Second example (for a first-order DFW): -- We have
presented above our experimental results for a dynamic FW of the
zeroth order. \ The most intriguing cases are, however, the ones
dealing with higher-order dynamic FWs. \ For the case of
higher-order (static) FWs, in previous papers we theoretically
constructed electromagnetic or acoustic {\em static envelopes}
with the shape of donuts, or of cylindrical
surfaces\cite{LWs,FW2,FW4,Patent}. \ Going on to higher-order
DFWs, striking results appear: We present in our last Figure the
experimental results referring to a first-order DFW, consisting in
a cylindrical surface of light {\em whose geometry changes in
space and in time.}

\h Our theoretical choice for a first-order DFW is as follows: We
may theoretically construct a higher-order DFW, characterized by a
duration $T$, whose behavior may be divided into three steps, with
a duration $T/3$ each, so that $T_1=T/3$ and $T_2=2T/3$; and with
a deformation speed (in each step) given by the alteration length
divided by $T/3$.  \ Namely, let us devise a FW initially
consisting in a short cylindrical surface of length $\Delta l_i$
and radius $\rho_i$, such that: \ (i) in the first step, the
cylindrical surface is required to stretch, keeping its radius,
till reaching the length $\Delta l_f$; \ (ii) in the second step,
the cylindrical surface keeps its length, but is requested to
reduce continuously its radius, till the value $\rho_f$; \ (iii)
in the third step, the cylindrical surface ---while keeping its
radius $\rho_f$--- is required to split into two cylinders, which
become shorter till reaching the length $\Delta l_f$ each.

\h Let us choose (for all steps) the simple function $F(z,t)$
defined as zero everywhere (within $0 < z < L$) except \bb
 F(z,t) \ug \left\{\begin{array}{clr}
 1 \;\;\; & {\rm for}\;\;\; z_1 < z < z_2  \\
 \\
 1 \;\;\; & {\rm for}\;\;\; z_3 < z < z_4 \; ,
\end{array} \right. \label{F}
 \ee
\noindent depending on time through the values of $z_1$, $z_2$,
$z_3$ and $z_4$. Its radius will change in time according to
$Q=Q(t)$.

\h To be more specific:

\h (i) {\em first movement,} $0 < t < T_1 \equiv T/3$: During this
phase, the cylinder radius $\rho_0 = \rho_i$ does not change, and
the corresponding value of $Q$ remains fixed, according
Eq.(\ref{Q2}). \ The z values are \ $z_1 = z_{1i} - v_1 t$; \ $z_2
= z_3 =  z_{1i} + \Delta l_i/2 \equiv  \bar{z}$; \ $z_4 = (z_{1i}
+ \Delta l_i) + v_1 t$, that is, $z_1$ and $z_4$ change in time,
while $z_2$ and $z_3$ remain fixed. \ The initial values of $z_1$
and $z_4$,  that we call $z_{1i}$ and $z_{4i} = z_{1i} + \Delta
l_i$, are still to be fixed; together with the final value
$z_{1f}$ of $z_1$, with $z_1(T_1) = z_{1f} < z_{1i}$, which also
determines the final value of $z_4$, so that $z_{4f} \equiv
z_4(T_1) = (z_{1i} + \Delta l_i) + (z_{1i} - z_{1f})$.  \ The
final length of the cylinder will be $\Delta l_f = z_{4f} -
z_{1f}$, while the speed $v_1$ will be $v_1 = (z_{1i} -
z_{1f})/T_1 = 3(z_{1i} - z_{1f})/T$.

\h (ii) {\em second movement,} $T_1 < t < T_2 \equiv 2T/3$: The z
values are the same assumed at time $t=T_1$ in the previous step,
that is, \ $z_1 = z_{1f}$; \ $z_2 = z_3 \equiv \bar{z}$; \ $z_4 =
z_{4f}$; \ but quantity $Q$ will vary according to Eq.(\ref{Q2}),
\ where \ $\rho_0(t) = \rho_i - v_2(t - T_1)$. \ The final value
of the radius, \ $\rho_0(T_2) \equiv \rho_f$, \ is still to be
fixed, our request being merely that $\rho_f < \rho_i$. \ The
speed $v_2$ will be \ $v_2 = (\rho_i - \rho_f)/(T_2 - T_1) =
3(\rho_i - \rho_f)/T$.

\h (iii) {\em third movement,} $T_2 < t < T_3 \equiv T$: Quantity
$Q$ keeps constant and equal to the final value it acquired in the
previous step. \ The z values will however change as follows: \
$z_1 = z_{1f}$; \ $z_2 = \bar{z} - v_3(t-T_2)$; \  $z_3 = \bar{z}
+ v_3(t-T_2)$; \ $z_4 = z_{4f}$. \ The final value of $z_2$ (and
of $z_3$) remains to be fixed, for instance through a choice for
the final length ($\Delta l_f$) of the two smaller cylinders: In
fact, it will be \ $z_2(T) = z_{1f} + \Delta l_f$; \ and, for the
final value of $z_3$, one will have \ $z_3(T) = z_{4f} - \Delta
l_f$. \ The speed is \ $v_3 = (\bar{z} - z_2(T))/(T_3-T_2) =
3(\bar{z} - z_2(T))/T$.

\h This specifies all what is needed for the experimental creation
of the devised higher-order (first order) DFW.

\h Its experimental production: --- The experimental procedure is
the one illustrated in the previous sections 2 and 3. \ For the
definition interval of the DFW, that is, $0 \leq z \leq L$, it has
been chosen the value $L = 40$ cm, being the other parameters
chosen as: $T=10$s, \ $z_{1i} = 0.18$ cm, \ $z_{1f} = 0.05$ cm, \
$\Delta l_i = 0.004$ cm, $\rho_i = 35 \ \mu$m and $\rho_f = 25 \
\mu$m. With this we have $v_1 =  0.039 \rm{m}/\rm{s} $, $v2 = 3
\rm{\mu m}/\rm{s}$  and $v3 = 0.03 \rm{m}/\rm{s}$. \ As to
Eq.(\ref{FW_caminhante}), the value $N=12$ has been adopted.

\begin{figure}[!ht]
                \centering
                \includegraphics[scale=0.25]{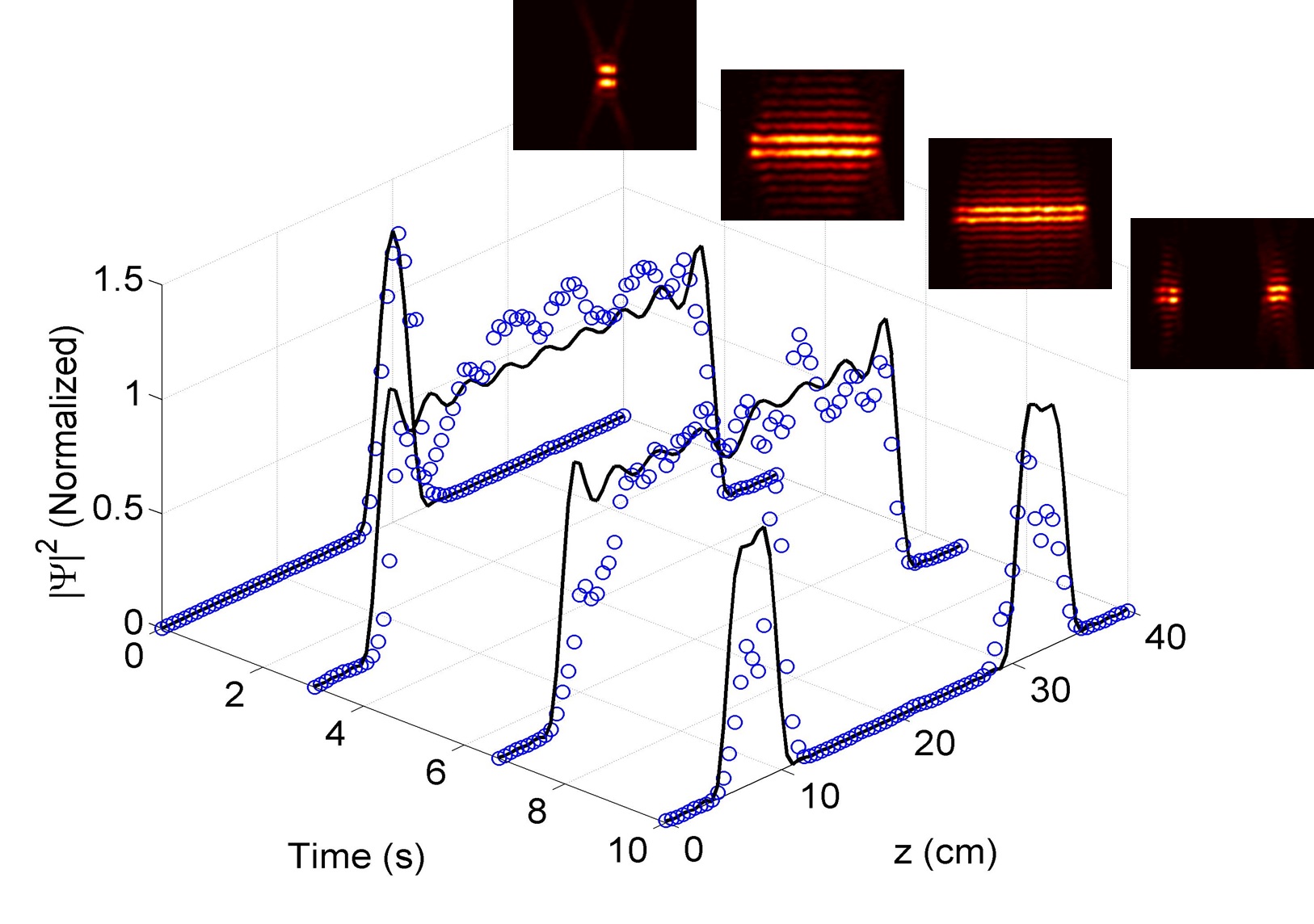}
                \caption{(color online) Time evolution, during $\Delta t = 10$ s, of the intensity pattern for our first-order dynamic FW; and orthogonal projections of the corresponding cylindrical surfaces: {\em experimental results.} \ The final radius of the cylindrical surface has been chosen to be \ $\rho_f = 25 \ \mu$, and its initial radius results to be \ $\rho_i = 35 \ \mu$. \ {\em Noticeably, as time elapses, the cylindrical surface stretches its length (while narrowing its radius), and afterwards splits into two shorter cylinders:} See also Media-2 online.}
                \label{Tarcio-ordem_1_com_proje}
        \end{figure}

\h In Figs.\ref{Tarcio-ordem_1_com_proje} the experimental results
are shown of the time evolution, during $\Delta t = 10$ s, of the
intensity pattern for our first-order dynamic FW; as well as the
orthogonal projections of the corresponding cylindrical surfaces.
\ The final radius of the cylindrical surface has been chosen to
be \ $\rho_f = 25 \ \mu$m, and its initial radius results to be \
$\rho_i = 35 \ \mu$m. \ {\em Noticeably, as time elapses, the
cylindrical surface stretches its length (while narrowing its
radius), and afterwards splits into two short cylinders:} see also
Media-2 online.

\h {\em 4. Conclusions} ---  The Non-Diffracting Waves (NDW) are
solutions to the (linear) wave equation which travel well confined
({\em localized\/}), in a single direction, and with other
interesting properties like self-healing. The Frozen Waves (FW)
are NDWs with zero peak-velocity, i.e. with a static envelope,
whose longitudinal intensity pattern can be freely modeled ---both
theoretically and experimentally---within a prefixed space
interval of the propagation axis. \ In this paper we present,
first, the theory of ``dynamic FWs" (DFW) whose shape is now
allowed to slightly evolve in time in a predetermined way, and,
second, their experimental generation by holographic techniques
(namely, by a holographic setup along with an optical
reconstruction of dynamic holograms CGHs by using a LC-SLM). The
most interesting experimental results, depicted in our last Figure
(and in movies online), refer to the higher-order FW consisting in
a cylindrical surface of light whose radius changes in space and
time. We show that, as time elapses, the cylindrical surface
stretches its length (while narrowing its radius), and afterwards
splits into two shorter cylinders. The experimental results are in
agreement with the theoretical predictions, and open exciting
possibilities of generating many further potentially interesting
DFWs for scientific and technological applications.


\end{document}